\documentclass[prc,twoside,floatfix,superscriptaddress,showkeys,showpacs,twocolumn]{revtex4}

\usepackage{graphicx, latexsym, amssymb, amsmath, color, multirow, mathrsfs, CJK, ifpdf}
\usepackage[section]{placeins}
\usepackage{amsmath}
\usepackage{amsfonts}
\usepackage{amssymb}
\usepackage{bm}
\usepackage{graphicx}
\usepackage{epstopdf}
\usepackage{color} 
\usepackage{txfonts}
\usepackage{ulem}

\makeatletter

\newcommand{\Rmnum}[1]{\expandafter\@slowromancap\romannumeral#1@}
\makeatother
\usepackage{dcolumn}   
\usepackage{physics}
\usepackage[breaklinks=false]{hyperref}
\begin{document}

\preprint{APS/123-QED}

\title{Projected shell model description of nuclear level density: \\I. Collective, pair-breaking, and multi-quasiparticle regimes in even-even nuclei}

\author{Jiaqi Wang}
\affiliation{School of Physics and Astronomy, Shanghai Jiao Tong
University, Shanghai 200240, China}
\author{Saumi Dutta}
\affiliation{School of Physics and Astronomy, Shanghai Jiao Tong
University, Shanghai 200240, China}
\author{Long-Jun Wang}
\affiliation{School of Physical Science and
Technology, Southwest University, Chongqing 400715, China}
\author{Yang Sun}\email{sunyang@sjtu.edu.cn  (correspondence)}
\affiliation{School of Physics and Astronomy, Shanghai Jiao Tong University, Shanghai
200240, China}

\date{\today}
\begin{abstract}

There are overwhelmingly experimental evidences indicating that excited nuclear states are dominated by quasiparticle (qp) excitations, which form many-body configurations with broken nucleon-pairs from different orbitals. By taking these multi-qp states as building blocks for shell-model basis, we propose a novel shell-model method for the calculation of nuclear level density (NLD) in deformed nuclei. The shell-model diagonalization with two-body residual interactions yields a large ensemble of {\it eigenstates} of angular momentum and parity. We demonstrate that NLD as a statistical quantity depends sensitively on the structure of deformed single-particle states. As the first example to introduce this method, we take a well-deformed rare-earth nucleus, $^{164}$Dy, for which NLD has been studied extensively by the Oslo method. By a quantitative comparison with discrete levels from spectroscopic measurements, we show that while the pronounced step-wise structure in the low-energy NLD curve can be understood as the collective excitation and nucleon-pair breaking, the exponential growth of levels in the higher-energy NLD can be described by the combination of the broken-pair states, subject to the Pauli Principle. According to the nature of NLD with increasing excitation, we divide the entire NLD curve into (1) collective regime, (2) pair-breaking regime, and (3) multi-qp regime. We discuss the formation mechanism and characteristic features of NLD for the three regimes. In addition, the parity dependence and angular-momentum dependence in NLD are investigated with a strong emphasis on the structure effect. 

\end{abstract}


\maketitle

\section{Introduction}\label{intro}

Nuclear level density (NLD), defined as the number of nuclear levels per unit energy interval, $\rho (E) = \Delta N / \Delta E$, is one of the basic properties of atomic nuclei. NLD is a crucial ingredient in nuclear reaction theories, in particular, in the calculation of thermonuclear reaction cross-sections of astrophysical interests, as well as in fission calculations. Moreover, NLDs are required as important inputs to estimate energy generation in stars or to determine nuclear abundances in various astrophysical processes using nucleosynthesis networks. For practical applications, they are also useful in reactor physics, nuclear medicine, and nuclear technological applications e.g., in the transmutation of radioactive nuclear waste. 

The study of NLD can be traced back to the attempt of Hans Bethe who, in his seminal work \cite{Bethe1936,Bethe1937}, derived a simple analytical formula by treating the nucleus as a gas of non-interacting fermions moving in equally spaced single-particle orbitals and applying a pure statistical partition function approach. Over the years, this simple thermodynamic approach has been subject to various corrections to include shell and pairing effects which gradually led to the derivation of constant temperature model \cite{Gilbert1965}, back-shifted Fermi-gas model \cite{Dilg1973}, generalized superfluid model \cite{Ignatyuk1993}, etc. Consistent attempts have been made throughout nearly the last three decades to adjust their parameters empirically to obtain reasonable agreement with the available experimental data \cite{ld_param_1,ld_param_2,ld_param_3,ld_param_4,ld_param_5,ld_param_6}.

Apart from the phenomenological models, microscopic models have also been developed for NLDs. Several combinatorial models exist in literature \cite{combinatorial_1,combinatorial_2,combinatorial_3, microscopic_1, microscopic_2, microscopic_3} among which Refs. \cite{microscopic_1, microscopic_2} have been widely accepted in practical applications. An exact shell model diagonalization is cumbersome due to the large dimension of the Hamiltonian matrix elements and in principle, is only possible in a physically truncated orbital space for relatively lighter nuclei or for nuclei in the vicinity of shell closure. For example, for several nuclei belonging to $sd$-shell, full shell-model diagonalization have been performed to derive NLDs \cite{sd-shell_full_diag_1, sd-shell_full_diag_2}. 
Alternative methods based on shell model concept, namely, Monte Carlo method \cite{smmc_1,smmc_2,smmc_3,smmc_4,smmc_5},  stochastic estimation method \cite{stochastic}, moments method \cite{moment_1, moments_2}, Lanczos method \cite{lanczos_1} etc. have been developed to avoid the computational complexity of full Hamiltonian diagonalization while trying to retain the physical essence of the full diagonalization procedure. However, most of these approaches, being in their early stages of development, have only limited applicability \cite{limited_model_applicability}. 

It is by far generally agreed that, with increasing excitation energy, level density curves grow as exponential functions, which behave as a straight line in a logarithm plot, connecting the NLD at the low-energy region of discrete states to that at the neutron separation energy obtained from neutron-resonance spacing data. In Ref.\cite{Alhassid2008}, Alhassid {\it et al.}  discussed the linear function $b_0+b_1 E_x$. The parameters $b_0$ and $b_1$ are sensitively isotope-dependent \cite{Guttormsen2015}. For the Oslo method, which is one of the most advanced methods for experimental determination of NLD (see discussions in Section \ref{Oslo}), it is critically important to know precisely where the anchor points of the line lie. The Shell Model Monte Carlo (SMMC) method \cite{smmc_1,smmc_2,smmc_3,smmc_4,smmc_5} is more advantageous in the high-energy region owing to its statistical aspects. For the low excitation region, the experimentally-available discrete levels can provide only limited information. Therefore, the determination of NLDs at the low-energy region before the exponential behavior sets in, where the NLDs are totally nuclear structure-sensitive \cite{Guttormsen2015}, is of great theoretical interest. It is purely a nuclear structure problem, for which the development of novel shell-model methods is required.

The present work is devoted to introducing a state-of-the-art shell-model method for the systematical calculation of nuclear level density and $\gamma$ strength function, with an emphasis on detailed structural information. We apply the idea of the Projected Shell Model by Hara and one of the present authors (YS) \cite{Sun2016,PSM-review,Sun1996PReport,PSMcode}, which has been proven to be successful in the description of nuclear spectroscopy for a wide range of isotopes of the nuclear chart, from the light nuclei \cite{Hara1999,Long2001} where full diagonalizations of large-scale shell-model are feasible, to the superheavy region \cite{Herzberg2006,Sun2008,Falih2009,Sun2010}, from which one may extract information for the anticipated superheavy island of stability. The present article is designed as the first one of a series of publications, in which we discuss NLD fully from a microscopic point of view, by connecting calculated NLD curves to detailed nuclear structure questions. We demonstrate that NLD as a statistical quantity depends sensitively on the structure of deformed single-particle states. 

The article is arranged as follows. As the results from the present theoretical method will be discussed with a close comparison with the experimental NLD results of the Oslo method, in Section II, we give a brief account on how the Oslo method generates NLDs from measured $\gamma$ rays. Section III discusses an important aspect of the basic nuclear structure in excited nuclear states, where the historic treatment of the nucleon pairing by the Bardeen-Cooper-Schrieffer (BCS) theory is reviewed. The discussion in this section emphasizes the unique physical process of pair breaking under nuclear rotation, for which we point out that in sharp contrast to sudden phase transition seen in other systems, the evolution of pair breaking in rotating nuclei is gradual over a wide range of excitations. Section IV outlines the Projected Shell Model, which is a novel shell model based on the deformed basis with the angular-momentum projection technique. In this section, we introduce the deformed single-particle basis, the multi-quasiparticle (qp) configuration space, the effective two-body Hamiltonian, and the state-of-the-art many-body method that allows us to carry out numerical calculations. Results and discussion with $^{164}$Dy as the example are given in Section V with four subsections. In Section V-A, we validate our model by showing a quantitative description of discrete nuclear levels in $^{164}$Dy, with a one-to-one comparison of our calculated energy levels with the known data. Based on these results, we present in Section V-B our theoretical NLD together with that of experimental discrete levels and the Oslo NLD curve. The focus of discussion in this subsection is the step structures in the NLD curve at the low-energy region of $^{164}$Dy. Our results suggest a missing step structure beyond 2.0 MeV of excitation in the Oslo curve, which corresponds to a simultaneous breaking of neutron- and a proton-pair. In a finite self-bound atomic nucleus, various classes of eigenstates characterized by different quantum numbers, such as total angular momentum $I$ and parity $\Pi$, can be very different. Section V-C discusses the structure-dependent behavior of even- and odd-parity NLDs at different excitations. Different from the usual thought, we show that for $^{164}$Dy, nearly doubled NLD of odd parity is predicted as compared to that of even parity. We provide structure explanations of why it is so. Furthermore, the distribution of NLD with different angular momenta (spins) is shown in Section V-D, where we find that for low-energy regions before the consummation of pair breaking, spin distribution patterns are generally irregular. However, with increasing energy beyond 4 MeV or so, a Gaussian-like distribution emerges, consistent with that of the statistical model assuming random coupling of angular momenta. As the present PSM and the previous SMMC approaches are both quantum many-body methods, which aim to study the same problem from very different viewpoints, in Section \ref{Compare-SMMC}, we have given a brief comparison of the two models. Finally, a summary of the work with the prospect of future works is given in Section \ref{Summary}.

\section{Experimental advances: The Oslo method}\label{Oslo}

During about last thirty years, the nuclear physics group at the Oslo Cyclotron Laboratory has been applying a novel experimental tool based on particle-$\gamma$ ray coincidence technique to simultaneously extract the nuclear level density and $\gamma$-ray strength function in the quasi-continuum and continuum regions of the nuclear excitation spectrum below the particle threshold \cite{oslo_0}. As many discussions in the present theoretical work are closely related to the Oslo NLD curve, we briefly recall the technique and the basic assumptions in its data analysis. 

In the Oslo method, a stable or long-lived target nucleus is bombarded with light ions to initiate a charged-particle direct reaction or inelastic scattering reaction, and $\gamma$ rays emitted from the excited daughter nucleus are detected in coincidence with the emitted ejectile. Next, a raw particle-$\gamma$ ray coincidence matrix is formed from which a primary $\gamma$-ray matrix is obtained after carefully selecting only first-generation $\gamma$ rays. After normalizing this primary matrix to unity, it is fitted to two one-dimensional functions of nuclear structure quantities, namely, nuclear level density and $\gamma$-ray strength function. 
In this step, a basic assumption is made that the decay probability is proportional to the NLD at the final energy following Fermi's golden rule \cite{fermi} and the decay is also proportional to the $\gamma$-ray transmission coefficient which is assumed to be independent of initial and final states as per the  Brink-Axel hypothesis \cite{brink,axel}.
Finally, the unique physical solutions of the two structure functions are obtained after determining their scale and slope parameters through normalization with known data. 

The Oslo method, together with its two recent parallel extensions, namely, the $\beta$-Oslo method \cite{beta-Oslo} and the inverse Oslo method \cite{inverse-Oslo}, has produced a wealth of nuclear level densities for stable as well as unstable radioactive nuclei over a wide range of excitation energy. However, the method does not provide NLDs as a function of spin and parity. 
Moreover, apart from the fact that uncertainties and systematic errors \cite{oslo_uncert2} occur due to the experimental limitations and  assumptions made at various steps of the method \cite{oslo_assumption}, the normalization procedure involves certain model dependencies which additionally gives rise to sizable uncertainties \cite{oslo_uncert1,oslo_uncert2}. 
Recently, as a model-independent approach, the \textit {shape method} \cite{shape_method} has been introduced which offers a universal and consistent prescription 
for determining the slope of the $\gamma$-ray strength function and the NLD function (when extracted simultaneously in the Oslo method) in the absence of experimental auxiliary data for normalization. However, one deficiency of the method is that it cannot provide the absolute values of the $\gamma$ strength functions when neutron resonance widths are not available from experimental measurements.

\section{Main structure of excited nuclear states }\label{shell}

The pairing correlation is an important aspect in nuclear structure and plays a decisive role in the discussion of NLDs \cite{Zelevinsky2019}.
In the ground state of a nucleus, the system tends to couple nucleons into pairs. The concept of nuclear superfluidity, analogous to that of electrons in condensed-matter physics described by the BCS theory \cite{BCS1957}, is crucial for explaining many basic properties in nuclei throughout the nuclear chart. In fact, nuclear physicists adopted the BCS theory into nuclear structure \cite{Bohr1958,Belyaev1959} immediately after its establishment in 1957. For example, the observed moment of inertia (MoI) in nuclei near their ground state could only be explained when pairing correlation is considered \cite{Ring2004many-body}. Soon after, it was realized that when collective rotation sets in, pair-breaking occurs due to the Coriolis anti-pairing effect \cite{MV1960}. Furthermore, thermal excitations provide energy to overcome the pair correlation (measured by the pair gap $\Delta$), which breaks the pairs. Thus generally, configurations of multi-qp states in the presence of pairing correlation dominate the structure of NLDs in excited regions of nuclei.

Nuclei, being small quantum systems with typically $10^1-10^2$ particles, are not ideal places to accomplish truly a phase transition between the superfluid state and the normal state. It was pointed out \cite{smmc_3} that the phase transition in the mean-field level is washed out via quantum fluctuation. The pairing transition itself has also been discussed theoretically and experimentally. Calculations showed \cite{Esebbag,Nakada} that the quantum effects can wash out or delay the phase transition. The situation is clearly distinguished from condensed-matter systems, mainly because in nuclei, nucleons sit in states labeled by different $j$'s, with $j$ being total single-particle angular momentum coupled by orbital angular momentum $l$ and spin $s\equiv 1/2$. When a nucleus is excited, nucleon pairs with higher-$j$ orbitals, due to the stronger Coriolis effect, are more easily broken. Hence, nuclear excitation is accompanied by {\it successive} processes of pair-breaking. One well-known example is the pair-breaking along the Yrast band (the rotational band consisting of the lowest energy levels for each spin). In this process, the observed rapid increase in MoI at spin $I\approx 10 - 14\hbar$ in rare earth nuclei is explained by the first band crossing between the ground-state band (g-band) and a 2-qp band (the so-called $S$ band \cite{SS1972}), where a pair of neutrons in the highest $j$-orbit ($j=i_{13/2}$) are broken. As the nucleus rotates faster and the Coriolis effect becomes stronger, a band crossing between 2-qp and 4-qp bands could account for the observed second increase of MoI at $I\approx 22\hbar$ \cite{Lee1977}, corresponding to the breaking of an additional proton pair in $j=h_{11/2}$. Another, and even more popular example is the wide existence of $K$-isomeric states in the nuclear chart, especially in those well-deformed nuclei with axial symmetry \cite{Walker1999}. The formation of $K$-isomers involves the breaking of more nucleon-pairs that align their individual angular momenta with the rotation \cite{Walker2016}. There are numerous experimental evidences (see, for example, Refs. \cite{Burde1982,Jensen2000,Yadav2009,Swati2022}) of long-lived $K$-isomeric states, and their formation has been interpreted as the breaking of three, four, or more nucleon pairs. These $K$-isomers are outstanding because of their suppressed transitions due to the selection rules. It is natural that many other non-isomeric excited nuclear states are formed in the same way by broken pairs.

These experimental evidences give us a clear hint for constructing shell-model configurations for NLD calculations. Accordingly, we build our many-particle configuration space in terms of multi-qp states \cite{Sun2016}. We may regard this classification of many-particle states as a generalization of the seniority concept \cite{Isacker2010}, which describes the degree of unpaired particles in the spherical basis, to deformed systems. The theoretical treatment of the quantum states with broken-pair configurations can be realized by the Tamm-Dancoff method \cite{Ring2004many-body}, in which the many-particle states are written as linear combinations of different orders of qp-configurations. For a deformed nucleus, it is desired to work with a deformed basis having a deformation closer to the `true' nuclear deformation \cite{Raman2001} (see also discussions in \cite{Sun2016}). Furthermore, such a deformed basis incorporates efficiently important correlations through the concept of spontaneous symmetry breaking  \cite{Sun2016}. The violated quantum numbers in the deformed basis can be recovered by the projection technique \cite{Guidry2022}. As we shall see below, the final shell-model diagonalization is then carried out in the (angular-momentum) projected basis
defined in the laboratory frame.

\section{Outline of the projected shell model}\label{PSM}

To describe deformed nuclei, it is convenient to adopt a deformed potential to start with, such as that of the Nilsson model that generates deformed single-particle states \cite{Nilsson1969}. In a deformed potential, the single-particle distribution near the Fermi surface is qualitatively different from that at the spherical limit. One then has to answer the question that how to efficiently treat the many-particle problems for deformed nuclear systems using the shell-model concept. 
The solution lies in the numerical angular momentum projection \cite{Guidry2022}, for which the calculation started in the 1970's \cite{Hara1979} and becomes popular today for nuclear structure calculations. One representative example applying this technique is the Projected Shell Model (PSM)
\cite{PSM-review,Sun1996PReport,PSMcode}. 

The PSM calculation starts from the deformed Nilsson model \cite{Nilsson1969} with pairing correlations incorporated from the BCS calculation \cite{PSM-review}. Three major harmonic oscillator shells with $N=4, 5, 6$ $(N=3, 4, 5)$ are considered to have orbitals for valence neutrons (protons) for rare earth nuclei. The standard Nilsson parameters ($\kappa$ and $\mu$) have been adopted from the literature \cite{Nil-1985,nilsson_param_1,nilsson_param_2,nilsson_param_3,Nil-1990} to generate deformed single-particle states for different mass regions. The PSM multi-qp model space is constructed from the deformed Nilsson+BCS qp basis. The multi-qp configurations up to $6$-qp states can be written as
\begin{equation}\label{shell_config}
\begin{aligned}
\phantom{A} &\phantom{=}\{\ket\phi, a_{\nu_{i}}^{\dagger}a_{\nu_{j}}^{\dagger}\ket\phi,
a_{{\pi}_{i}}^{\dagger}a_{{\pi}_{j}}^{\dagger}\ket\phi, a_{\nu_{i}}^{\dagger}a_{\nu_{j}}^{\dagger}a_{\pi_{k}}^{\dagger}a_{{\pi}_{l}}^{\dagger}\ket\phi, \\
&   a_{\nu_{i}}^{\dagger}a_{\nu_{j}}^{\dagger}a_{\nu_{k}}^{\dagger}a_{{\nu}_{l}}^{\dagger}\ket\phi,
  a_{\pi_{i}}^{\dagger}a_{\pi_{j}}^{\dagger}a_{\pi_{k}}^{\dagger}a_{{\pi}_{l}}^{\dagger}\ket\phi, \\
& a_{\nu_{i}}^{\dagger}a_{\nu_{j}}^{\dagger}a_{\nu_{k}}^{\dagger}a_{{\nu}_{l}}^{\dagger}a_{\nu_{m}}^{\dagger}a_{\nu_{n}}^{\dagger}\ket\phi,   a_{\pi_{i}}^{\dagger}a_{\pi_{j}}^{\dagger}a_{\pi_{k}}^{\dagger}a_{{\pi}_{l}}^{\dagger}a_{\pi_{m}}^{\dagger}a_{\pi_{n}}^{\dagger}\ket\phi, \\ 
& a_{\pi_{i}}^{\dagger}a_{\pi_{j}}^{\dagger}a_{\nu_{k}}^{\dagger}a_{{\nu}_{l}}^{\dagger}a_{\nu_{m}}^{\dagger}a_{\nu_{n}}^{\dagger}\ket\phi, a_{\nu_{i}}^{\dagger}a_{\nu_{j}}^{\dagger}a_{\pi_{k}}^{\dagger}a_{{\pi}_{l}}^{\dagger}a_{\pi_{m}}^{\dagger}a_{\pi_{n}}^{\dagger}\ket\phi, ~\dots\}.\\ 
\end{aligned}
\end{equation}
where, $a_{\nu}^{\dagger} (a_{\pi}^{\dagger})$ labels neutron (proton) qp-creation operator associated with the deformed qp-vacuum $\ket\phi$. The ``$\dots$" in (Eq.~\ref{shell_config}) denote those higher order qp states that are not included in the present calculation but can be added back when necessary. As the PSM works with multiple harmonic-oscillator shells for both neutrons and protons, the indices $\nu$ and $\pi$ in Eq.~\ref{shell_config} are general. For 2-qp configurations, for example, $a_{\nu_{i}}^{\dagger}a_{\nu_{j}}^{\dagger}\ket\phi$ ($a_{{\pi}_{i}}^{\dagger}a_{{\pi}_{j}}^{\dagger}\ket\phi$) are built by considering all possible combinations of two neutrons (protons) from a same shell for positive parity (or even-parity, with the parity quantum number $+1$), or each from neighboring shells for negative parity (or odd-parity, with the parity quantum number $-1$) states, with the respective indices running over all possibilities from the valence shells. In the construction of the combinational states, the Pauli Exclusion Principle is fully considered.

The deformed multi-qp basis states are projected onto good angular momenta to form shell-model configurations in the laboratory frame. Finally the two-body shell-model Hamiltonian (see below) is diagonalized in the projected space which mixes different configurations. 
We would like to mention that the so-constructed qp configurations, labeled by the quantum numbers with the Nilsson notation \cite{Nilsson1969}, can usually find their experimental correspondence. For example, one could find the dominant 2-qp configurations that cause the observed back-bending effect in MoI of the Yrast rotational band in $^{48}$Cr \cite{Hara1999}, or the qp structure of the experimentally-identified 4-qp $K$-isomer in the superheavy element $^{254}$No \cite{Herzberg2006}. 


Our PSM Hamiltonian consists of separable forces:
\begin{equation}\label{two-body}
\hat{H}=\hat{H_{0}}-\frac{1}{2}\chi_{QQ}\sum_\mu \hat{Q}_{2\mu}^{\dagger} \hat{Q}_{2\mu}-G_{M}\hat{P}^{\dagger}\hat{P}-G_{Q}\sum_{\mu}\hat{P}_{2\mu}^{\dagger}\hat{P}_{2\mu},
\end{equation}
where, $\hat{H_{0}}$ is the spherical single-particle term including the
spin-orbit force \cite{Nil-1985}. 
The rest terms are quadrupole-quadrupole interaction, monopole-pairing interaction, and quadrupole-pairing interaction, respectively. It has been suggested by Dufour and Zuker \cite{Dufour1996} that these terms are essential in any realistic effective interactions to describe nuclear structure properly. 
The two pairing forces in (\ref{two-body}) are assumed to be of the isovector type and act only between the like nucleons. The coupling constant for the monopole-pairing force is taken to be of the following form;
\begin{equation}
G _{M} = \left(G_{1} \mp G_{2} \frac{N - Z}{A}\right)\frac{1}{A} ~\textnormal {in MeV},
\end{equation}
where, ``+" (``$-$") denotes protons (neutrons),
and $N$, $Z$, and $A$ are the neutron number, proton number, and mass number, respectively. The values of the coupling constants $G_{1}$ and $G_{2}$ are adjusted to reproduce the experimental odd-even mass differences in the respective nuclear mass region. These two values are fixed as $21.24$ and $13.86$, respectively, in the present paper. The quadrupole pairing force $G_{Q}$ is considered as proportional to $G_{M}$ by an overall factor of $0.18$. 

The numerical calculation of angular momentum projection involves the evaluation of the rotated matrix elements, which is the most technically challenging and time-consuming part of PSM calculations. For those involving higher orders of qp states, a breakthrough in computational many-body techniques is needed. In nuclear structure physics, the Pfaffian concept has been introduced by Robledo as a key mathematical tool for solving the long-standing problem in the phase determination of the rotated matrix elements \cite{Robledo2009}. Soon after its introduction, it has been realized that the Pfaffian algorithm is very efficient also for calculating overlap matrix elements \cite{Mizusaki2013,Hu2014}. The Pfaffian algorithm has been used to enlarge the PSM space to high-order qp configurations \cite{LJWang2014}, which enables an aggressive extension of qp configurations in the PSM up to 10-qp states \cite{LJWang2016}. A large number of energy levels (in the present work, $\sim 10^5$ eigenstates of angular momentum and parity) can be obtained with a reasonable computational effort. Based on the new development, a PSM analysis of structural evolution and study of chaoticity in fast-rotating nuclei was given in Ref. \cite{Wang2019}, and detailed rotational bands up to high spins could be obtained from one diagonalization \cite{Wu2017a,Wu2017b}.

\section{RESULTS AND DISCUSSION}

In the determination of level densities, it is crucial to get information on the true spin-parity distributions as functions of excitation energy, $\rho(E,I,\Pi)$. However, for a calculation of reaction rates, only a few experimental data are available that can provide detailed level densities with true spin-parity. As discussed above, we provide a novel shell-model method for arbitrarily heavy, deformed nuclei to  obtain discrete levels, with corresponding wavefunctions being eigenstates of angular momentum and parity. From these wavefunctions, observables such as electromagnetic transitions, decay rates, and other quantities related to thermodynamic properties, can be calculated. $\gamma$-rays of multipolarity are obtained directly for any given initial and final states.
\begin{figure*}
   \includegraphics[height=4in, width=7.75in]{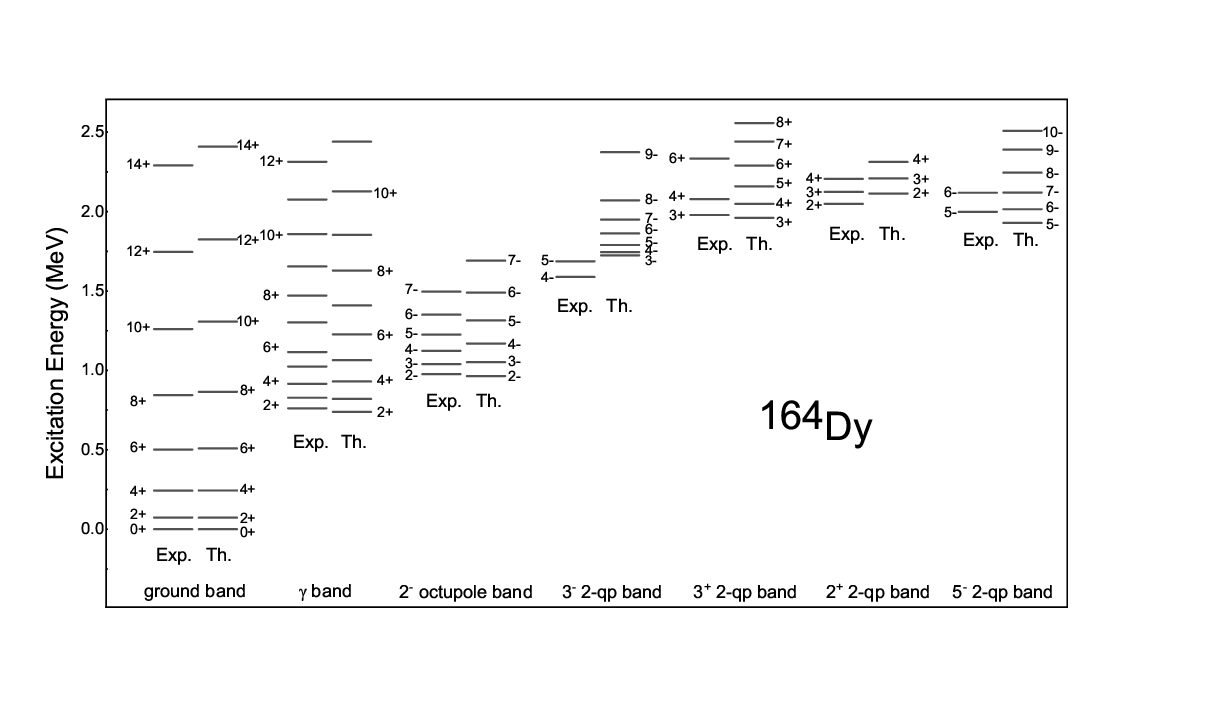} 
   \caption{\label{energylevels} Calculated energy levels for $^{164}$Dy, and comparison with available discrete levels taken from \cite{database} for the low-lying excited bands with known spin and parity.}
 \end{figure*}
In this section, we discuss a numerical example from rare-earth isotopes, for which we have sufficient knowledge about the deformation property. Nuclear deformation parameters can be extracted from experiment \cite{Raman2001}, or from systematical calculations with the finite-range droplet macroscopic and folded-Yukawa single-particle microscopic structure model \cite {Moller2016ADNDT}. We adopt the deformation parameters as inputs for our PSM calculation. 

\subsection{Quantitative description of discrete nuclear levels}

To describe NLD at a quantitative level, it is desired that the model is able to reproduce all experimentally-known discrete levels that are available in the published database. Without this step, one may doubt the model's predictive ability for cases where no data exist. For comparison of our calculation with the discrete nuclear data, we refer to the database from the National Nuclear Data Center at the Brookhaven National Laboratory \cite{database}. 

As one will see, the following discussion is mainly of a nuclear structure problem. For well-deformed even-even nuclei of the rare earth region, the structure of the low-energy NLD below 1 MeV is rather simple. For $^{164}$Dy, for example, there are only a total of eight excited states known in the database falling into the 1-MeV energy interval. They all belong to collective excitations; among which, the lowest three levels contributing to the NLD up to about 0.7 MeV of excitation are from the $I^\pi = 2^+, 4^+$ and $6^+$ states of the ground-state rotational band. The second known collective band starting from the $2^+$ state at 762 keV, which is followed by the $3^+$ state at 828 keV and $4^+$ state at 916 keV, belongs to the collective $\gamma$ vibrational band. Starting from somewhat higher energy at 977 keV, an odd-parity $2^-$ collective band is experimentally known to be an octupole vibrational band \cite{vibration}. Roughly from 1.5 MeV up, states with the pair-breaking dominate the structure. Thus, a correct theoretical description requires that the model can reproduce all these nuclear states.

The calculation is performed by using the PSM, outlined in Sect. \ref{PSM}, with its extensions. The extended versions of the PSM include the one with multi-qp configurations in the PSM up to 10-qp states \cite{LJWang2016}, the one with the triaxially deformed basis which enables a simultaneous description of $\gamma$ vibrational band \cite{gamma0,gamma1,gamma2}, and the one that spontaneously breaks reflection symmetry and can therefore describe low-lying octupole vibrational bands \cite{octupole1,octupole2}. The quadrupole deformation parameter for the $^{164}$Dy calculation is taken as $\varepsilon_2=0.28$. In Fig. \ref{energylevels}, we present partial energy levels calculated for $^{164}$Dy and compare them with the available discrete levels with known spin and parity, taken in \cite{database}. Overall, the experimental levels are well reproduced. For the first three collective bands, except for the high-spin states in the $\gamma$ band and in the $2^-$ octupole vibrational band where our calculated energies are higher than data, the other states are correctly described. The rest four experimentally-known bands are 2-qp bands with either even or odd parity which start from an excitation between 1.5 to 2 MeV. It can be seen from Fig. \ref{energylevels}, our calculation reproduces all of them satisfactorily. 

The overall good description for the low-lying states in the present  $^{164}$Dy example, as well as from many previous examples documented in the literature, suggests that the PSM (with its extensions) is a practical shell model for heavy, deformed nuclei. It is the promising model that can quantitatively describe the known discrete nuclear levels that the NLD models and the Oslo experiment often use for calibration of the low-energy side. However, as we shall point out in the following discussion that for many cases, the existing information extracted from the discrete level data must be taken with great caution.

\subsection{Comparison of NLD with discrete levels and the Oslo curve}

Before starting the discussion, we note that the currently-available discrete nuclear levels from the open database (such as those in the NNDC database \cite{database}) are generally far from being complete. Especially for those with excitation energy higher than 2 MeV, i.e. above the energy that is needed to break the first nucleon pair, discrete levels can not provide a complete picture for NLD. Furthermore, for many measured energy levels, spin and/or parity are uncertain. Therefore, great caution must be taken when using the information from the discrete levels to extrapolate the level density further up to connect that at the neutron separation energy derived from neutron resonance level spacing data.

While comparing our results with other theoretical models and with experimental data, we have used the definition of nuclear level density using the concept of energy bins. We remark that this is not a necessary step for presenting theoretical results, especially for the low-energy region with discrete levels where statistics does not make much sense. As our shell model produces individual nuclear state with definite spin and parity, we obtain level density $\rho(E)$ not as a smooth function of the excitation energy $E$, but rather as a sum of $\delta$-functions, each corresponding to one state with good spin and parity. Nevertheless, to present our results, we take roughly eight bins in a 1-MeV interval, with 0.12-MeV steps, consistent with the Oslo curves. In this way, our NLD curves are drawn with about eight data points in a MeV interval, with each point containing all levels counted in the 0.12-MeV step. As our NLD plots are given as usual in the unit of MeV$^{-1}$, a factor of $1/0.12 \approx 8.33$ is applied to our theoretical data points.

For the known discrete levels in $^{164}$Dy with experimentally-assigned spin, except for the first two collective bands, namely the ground-state band and the $2^+$ $\gamma$-band starting from 762 keV, where the measurement was extended to high-spin states, spin quantum numbers seldom exceed 10$\hbar$. Therefore, in our comparison with data, we include the calculated levels with spin $I\le 10\hbar$ only. With this spin cut-off, the comparison is made for level energies up to 4.5 MeV. Extensions to higher-spin states and/or to higher excitations are possible. However, further extension of our configuration space is rather time-consuming in computation. To compare with the Oslo curve, the experimental level density curve from the discrete levels is drawn to contain all levels (of known or unknown parity), and in our theoretical results, unless specified, we add levels with both parities together. Therefore, the NLD presented from our theoretical model calculation in this paper designates the following:
\begin{equation}
\rho(E) = \sum_{\Pi}\sum_{I ~\le ~10} \rho(E, I, \Pi).
\end{equation}

\begin{figure}\centering
\includegraphics[width=3.8in]{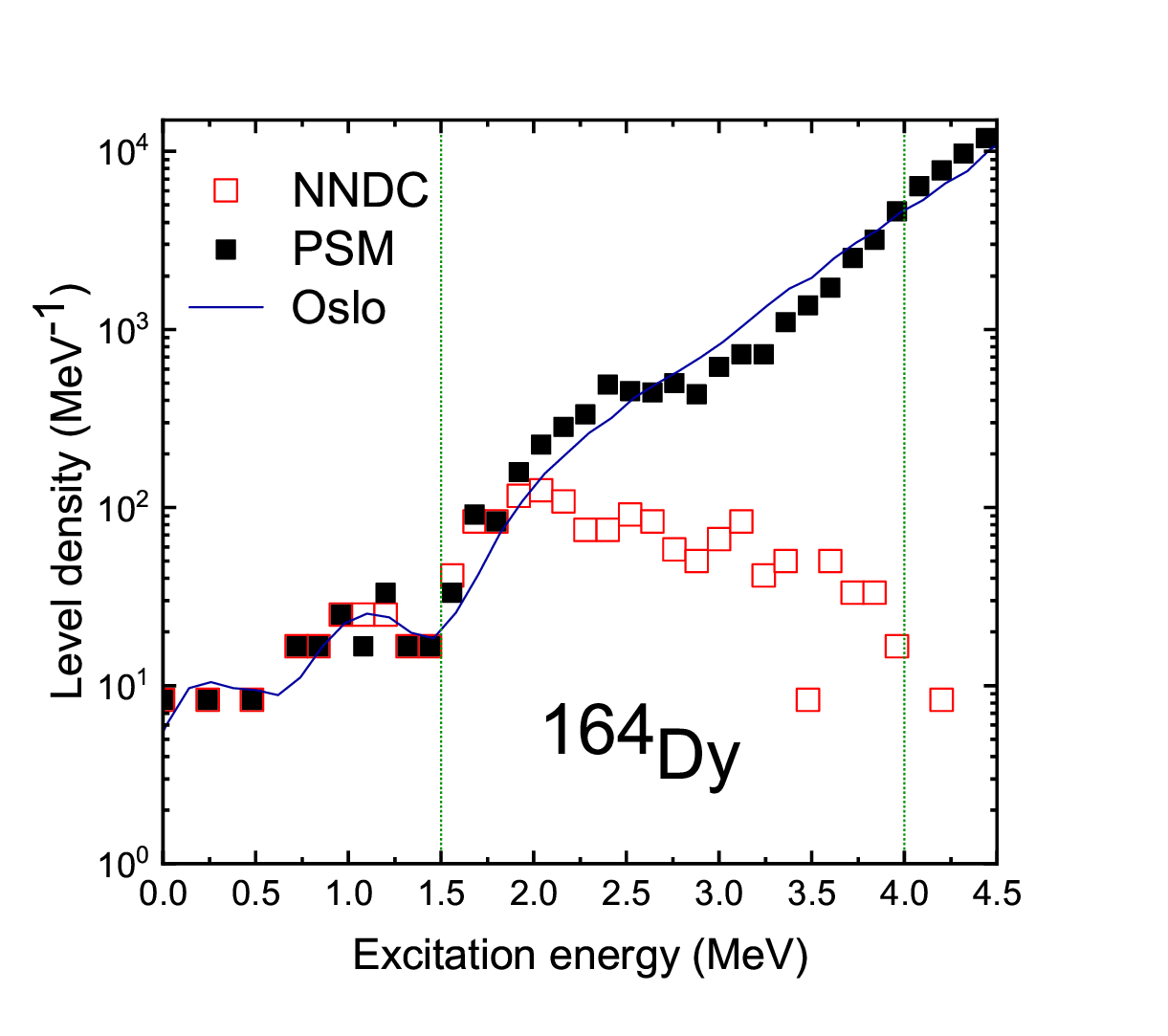}
\caption{\label{fig:level}(Color online) Calculated level density (filled squares) of $^{164}$Dy are compared with (1) experimental
level density obtained from the known discrete levels (red open squares) \cite{database}, and (2) level density of the Oslo method (blue solid line) \cite{Oslo1}.}
\end{figure}

In Fig. \ref{fig:level}, we show the calculated level density points (filled squares) for $^{164}$Dy, together with the experimental ones obtained from the known discrete levels (red open squares) \cite{database} and the level density curve of the Oslo method (blue solid line) taken from Refs. \cite{Oslo1}. All the curves are consistently given in 0.12-MeV energy bins. As one can see, a nearly perfect one-to-one match between our calculation and experimental discrete levels is found up to the peak of the curve from discrete levels at 1.75 MeV. The nice agreement of our calculation with the known discrete levels up to 1.75 MeV suggests that the model works correctly for describing the collective states in the low-energy regime (0 - 1.5 MeV range, see Fig. \ref{fig:level}) and the pair-breaking regime with some pair-breaking configurations contained in the bin energies larger than 1.5 MeV.

It is interesting and important to discuss step structures of the NLD curve at the low-energy region of $^{164}$Dy. Data of the discrete levels suggest noticeably two rapid rises in the curve, starting at the 0.75 and 1.5-MeV bin-energy, respectively. The former is understood as the contribution from the collective vibrational states and the latter corresponds to the beginning of the first pair-breakings following which states of qp excitations start to contribute to the NLD curve. The Oslo curve agrees nicely with the two step structures found in the discrete levels in $^{164}$Dy. 

As one can see from Fig. \ref{fig:level}, our theoretical curve agrees perfectly with both curves up to the 2.0-MeV bin, including both rises. Beyond 2.0 MeV, our calculated curve is predicted to keep rising, showing remarkably a third step structure. This structure is found to cross over a wide energy range from 2.3 to 4.0-MeV (see Fig. \ref{fig:level}). Within 2.4 - 2.9 MeV, a plateau-like structure shows up in the level density curve, indicating a nearly constant value in this energy interval. It follows by a rapid rise until it meets the Oslo curve at 4 MeV.

The appearance of the third step structure in NLD for $^{164}$Dy is at odds with the Oslo curve, which suggests a straight line in the logarithm plot from 2 MeV and up. The formation of our predicted third step structure is understood as follows. The second rise due to the first pair-breaking is seen to saturate in the 2.4 - 2.9 MeV plateau, meaning that in the energy range of the plateau, no new 2-qp states from one-pair breaking contribute to this energy interval. To keep the NLD curve rising, more nucleon pairs need to be broken to form new configurations. We find that the third rise is attributed to the beginning of the contribution from 4-qp states, which correspond mainly to a {\it simultaneous} breaking of two nucleon-pairs, a neutron- and a proton-pair. The required energy for breaking a neutron- and a proton-pair is $2\Delta_n$ and $2\Delta_p$, respectively, and therefore, the minimum energy for the 4-qp states is $2(\Delta_n+\Delta_p)$, which is about 3 MeV in the present case. As many such 4-qp configurations contribute successively to the NLD, we expect that the third rise starts at the 3.0-MeV bin and continues to 4.0 MeV and further up where the theoretical curve meets the Oslo one.

As seen in Fig. \ref{fig:level}, the experimental level density curve constructed from the known discrete levels (red open squares) begins to drop at the 2.0-MeV energy bin, which is about the energy needed to break one nucleon pair to form 2-qp states. This implies that a large amount of 2-qp and almost all the 4-qp configurations are missing in the spectroscopic data. The 4-qp state with breaking of one neutron- and one proton-pair in rare earth nuclei was experimentally confirmed a long time ago for the Yrast line \cite{Lee1977}. However, the present discussion talks about many excited 4-qp states above the Yrast line. The Oslo curve, which makes use of experimental discrete levels to determine the anchor point of the power-law curve, misses this important structure.

Based on the above discussions, we may divide the entire NLD curve into three regimes emphasizing their structures: the collective regime, the pair-breaking regime, and the multi-quasiparticle regime. In Fig. \ref{fig:level}, we draw vertical lines (dotted green lines) at 1.5 MeV and 4.0 MeV to separate the three regimes for $^{164}$Dy. The structure of the collective regime (0 - 1.5 MeV) is relatively simple. As the NLD in this regime is made by only a few well-known collective bands, it is not difficult to identify all the levels in this regime. A more pronounced structure effect occurs in NLD in the pair-breaking regime (1.5 - 4.0 MeV), which is characterized by the interplay between the collective (with all nucleons paired) and the pair-breaking configurations \cite{Wang2020}. The processes of one neutron- {\it or} one proton-pair breaking as well as the simultaneous breaking of a neutron- {\it and} a proton-pair happen in this regime. It is to be noted that because of the isospin degree of freedom, the formation of two kinds of Fermionic pairs, neutron-pairs and proton-pairs, and the existence of respective pairing gaps, $\Delta_n$ and $\Delta_p$, are unique properties only in nuclear systems. 
 Our inference is that the appearance of the additional step structure in NLD at about $2(\Delta_n+\Delta_p)$ in energy is a manifestation of two distinct Fermionic pair gaps in nuclei. Finally, with excitation energy going further up, the system enters into the multi-qp regime beyond 4.0 MeV, where the NLD approaches the order of $10^4$ in even-even nuclei (see Fig. \ref{fig:level}). As more and more nucleon pairs are broken, the system evolves from the state of collective motion into a state characterized by chaotic motion \cite{Weidenmueller2009,Gomez2011}. Features of the dynamics in a chaotic nuclear system are controlled by residual interactions of quasi-particle configurations. The actual stationary states become extremely complicated superpositions of the original simple configurations. Zelevinsky {\it et al.} call this process {\it stochastization} \cite{sd-shell_full_diag_1}. For this reason, the multi-qp regime can also be called the chaotic regime. With increasing complexity in this regime, the characteristic individual configurations are smoothed out completely. The NLD curve begins to follow the power law, as predicted by the constant temperature model.

Our finding in the pair-breaking region revises the general belief that the exponential behavior of the NLD sets in immediately when the first nucleon pairs are broken at $E> 2\Delta$ \cite{Guttormsen2015} (which implicitly assumed the same $\Delta$ for neutrons and protons). Our calculation generally suggests a much delayed observation of the exponential behavior in NLD, due to the recognition of the simultaneous breaking of a neutron- and a proton-pair. This implies that structure dominance by individual configurations becomes more pronounced at the collective and pair-breaking regimes. In the present $^{164}$Dy example, our calculated NLD curve begins to follow the power-law beyond 4 MeV where it meets with the Oslo curve (see Fig. \ref{fig:level}).

\subsection{Structure dependence of NLD for opposite parities}

In nuclei, the state of each nucleon has even or odd parity, depending on its orbital angular momentum $l$. The parity of many-nucleon configurations can be predicted in the present PSM calculation. For a given excitation energy, it is usually thought that the total nuclear states, thus NLD, can be divided into two equal groups of even and odd parity \cite{Ericson1960}. However, this conclusion was obtained from a statistical consideration by totally neglecting the shell effect. Our following discussion suggests that it is generally not the case. Depending on the structure, the NLD for decomposed parities at a given excitation energy may differ considerably. We take the present example of $^{164}$Dy to show that for the discussed energy region up to 4.5 MeV, the ratio of the level density for opposite parities fluctuates with different excitation energies. Below, we explain in detail why this can happen.

\begin{figure}
\centering
\includegraphics[width=3.7in]{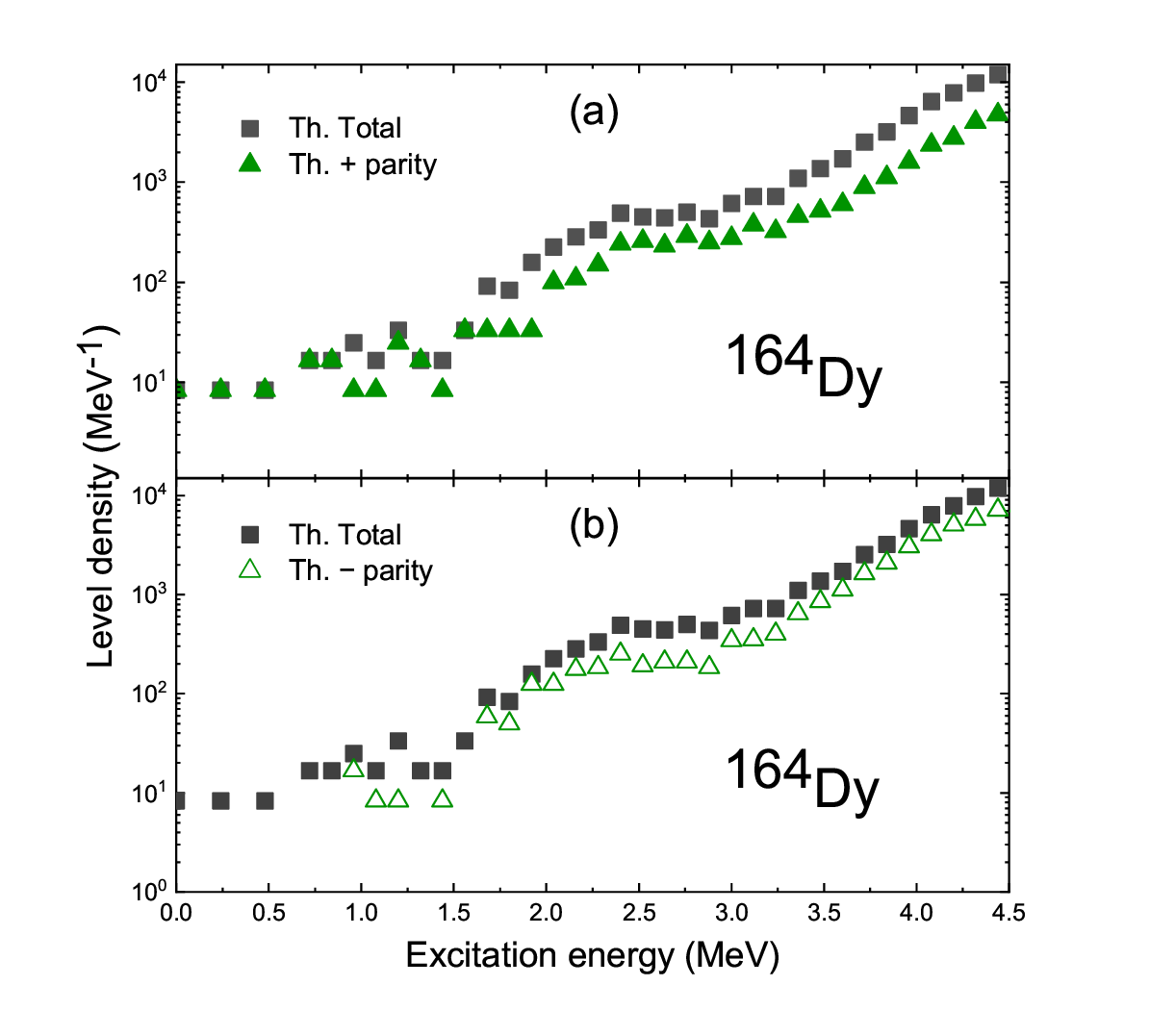}
\caption{\label{fig:parity}(Color online) Calculated total level density in $^{164}$Dy (filled squares, taken from Fig. \ref{fig:level}) are separately shown in (a) $+$ parity (filled triangles) and (b) $-$ parity (open triangles).}
\end{figure}

\begin{figure}
\centering
\includegraphics[width=3.7in]{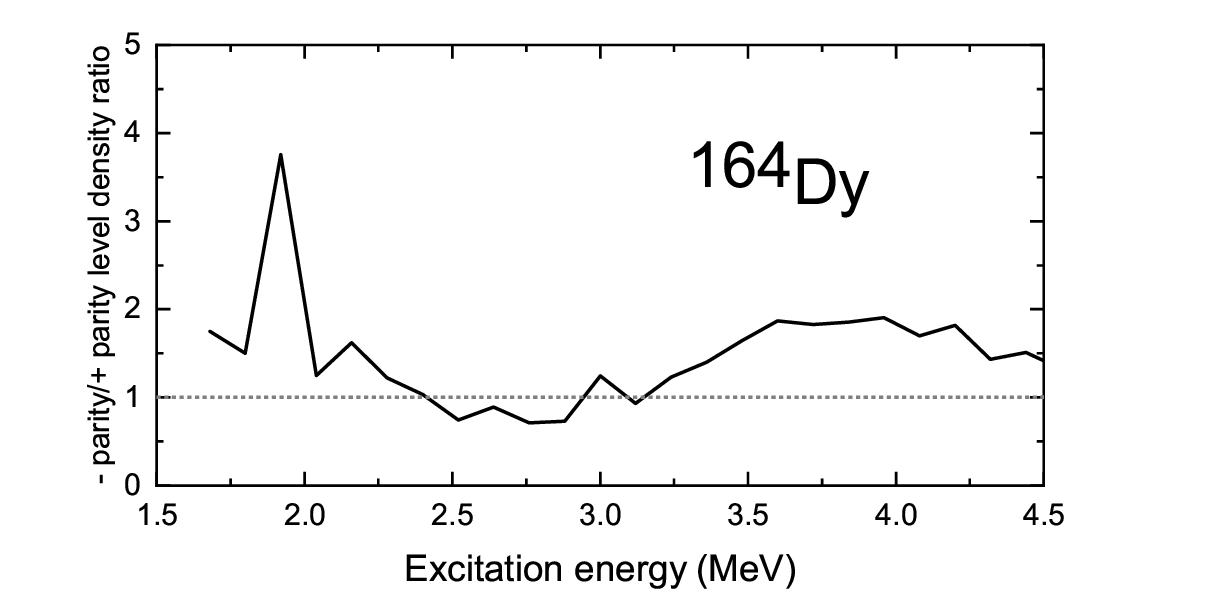}
\caption{\label{fig:parity-ratio} Ratio of calculated level density of $-$ parity and that of $+$ parity. The numbers are taken from Fig. \ref{fig:parity}.}
\end{figure}

By the symmetry requirement, in even-even nuclei, the energy levels in the ground-state band and the two common types of collective vibrational bands (i.e. $\beta$- and $\gamma$-vibrational bands) have even parity. In contrast, odd-parity states occur in collective octupole vibrational bands; however, the formation must meet the special requirement in structure. The condition for the occurrence of low-lying octupole collectivity in nuclei is that the deformed single-particle orbitals near the respective Fermi surfaces consist mainly of those that satisfy $\Delta l = 3$ \cite{Ahmad1993}. In the present $^{164}$Dy example, the proton Fermi surface lies in the vicinity of $h_{11/2}$ and $d_{5/2}$, and the neutron Fermi surface is close to both $i_{13/2}$ and $f_{7/2}$. In fact, a $2^-$ octupole vibrational band is experimentally observed in $^{164}$Dy at very-low energy starting at 977 keV \cite{database}, thus contributing to the odd-parity states of the first NLD regime.

On the other hand, odd-parity states in deformed nuclei can appear as qp states in the second and third regimes. In even-even nuclei, for example, odd-parity 2-qp states can be formed by two (broken-pair) nucleons sitting separately in the neighboring major shells having opposite parity. We find that in $^{164}$Dy, there are many orbitals available to contribute to odd-parity configurations. They include the neutron single-particle orbitals $\nu \frac{7}{2}^+[633]$, $\nu \frac{1}{2}^-[521]$, and $\nu \frac{5}{2}^-[512]$ above the $N=98$ gap, and $\nu \frac{5}{2}^+[642]$ and $\nu \frac{5}{2}^-[523]$ below it. One also finds the proton single-particle orbitals $\pi \frac{7}{2}^-[523]$ and $\pi \frac{1}{2}^+[411]$ above the $Z=66$ gap, and $\pi \frac{3}{2}^+[411]$ and $\pi \frac{5}{2}^+[413]$ below it. Because the $^{164}$Dy neutron- and proton-Fermi levels both lie inside of their respective energy gap in the deformed Nilsson states, abundant odd-parity 2-qp states can be formed at very low excitations, ranging from 1.6 to 2.4 MeV (see the following discussions).

The fact, that odd-parity 2-qp states occur more in number than their even-parity counterpart, can be recognized in $^{164}$Dy NLD curves if we plot states with opposite parities separately. In Figs. \ref{fig:parity}(a) and (b), we show theoretical NLD curves drawn separately for even and odd parities and compare each of them with that of the total NLD curve seen in Fig. \ref{fig:level}. Both plots, as in Fig. \ref{fig:level}, include all calculated spin states up to $I = 10\hbar$, i.e.
\begin{equation}
\rho(E, \Pi) = \sum_{I ~\le ~10} \rho(E, I, \Pi).
\end{equation}
It can be recognized from Figs. \ref{fig:parity}(a) and (b) that for most of the bin-energies, our results clearly indicate a non-half-half allocation for opposite parities. In the pair-breaking and multi-quasiparticle regimes, the number of levels for a given bin-energy is obviously different for different parities, depending on the structure of deformed single-particles. To see the differences clearly, we draw in Fig. \ref{fig:parity-ratio}, the ratio of the NLD curves of odd and even parities. The first and sharp peak in the ratio is seen at 1.9 MeV, where the number of odd-parity states is more than that of even-parity by nearly a factor of four. The second and wider peak is extended over the energy range from 3.2 to 4.5 MeV. The appearance of the second peak is attributed to the odd-parity 4-qp states formed by the same odd-parity 2-qp states which make the first peak plus a pair of 2-qp states of even parity. Some studies exist in the literature which also reported that the basic assumption of equal parity distribution for NLDs were often not realized. For example, in Ref.~\cite{parity_ratio_mocelj}, authors investigated the parity distribution for several nuclei in the Fe region and concluded that the equal parity ratio was not fulfilled at low excitation energies, not even at the particle separation energies. In Ref. \cite{Alhassid_00}, from shell model Monte Carlo calculation as well as by using a simple formula, Alhassid {\it et al.} found that at low energies, only a single parity dominated and cross-over to equal parity ratio were seen only at sufficiently higher excitation energies. Early theoretical work based on combinatorial model \cite{Herman_87,Herman_88,Cerf_91_combinatorial} and equidistant model \cite{Cerf_92_equidistant} also found deviations from equiparity approximation.

We remark that our $^{164}$Dy results in Figs. \ref{fig:parity} and \ref{fig:parity-ratio} demonstrate sensitive dependence of the parities on the structure which should not be taken as a general conclusion for the allocation of the parity states. As we have seen, NLDs are generally determined by the detailed structure of a nucleus under discussion, especially at the pair-breaking regime. In Figs. \ref{fig:parity}(a) and (b), one sees similar low-energy oscillations and the plateau structure in the 2.4 - 2.9 MeV energy range in the decomposed parity curves, suggesting that the structure effect in a given nucleus totally determines the NLD behavior. Nevertheless, from our present findings, we conclude that the sensitive structure dependence on parity in the NLD curve is another important consequence of the present shell-model study.

\subsection{Spin distribution in NLD}

When applying NLD in nuclear reaction calculations, for instance in the study of thermonuclear reaction cross-sections of astrophysical interests, information on NLD with definite angular momentum (i.e. spin) states is needed. 
In early works by Ericson \cite{Ericson1959,Ericson1960}, a formula to estimate spin-distribution of NLD, $\rho(E,I)$, was first introduced by using a statistical model assuming random coupling of angular momenta. Ericson obtained a Gaussian-like distribution (see also Ref. \cite{Larsen2019}) 
\begin{equation}\label{spin-distribution}
\rho(E, I) \approx \rho(E) \frac{2I+1}{2{\sqrt{2\pi}}\sigma^3} \exp{-\frac{I(I+1)}{2\sigma^2}},
\end{equation}
where $\rho(E)$ is the spin-independent level density. The actual shape of the distribution is determined by dispersion $\sigma$, which is an unknown parameter in the model \cite{Ericson1959}. Ericson argued that if the nuclear moment of inertia, ${\mathcal{J}}$, can be approximated by the classical rigid value, ${\mathcal{J}}_{\rm rigid}$, then a relation 
\begin{equation}\label{sigma}
\sigma^2 = \frac{\mathcal{J} T}{\hbar^2}
\end{equation}
holds \cite{Ericson1960}. The above relation involves a thermodynamic quantity, the nuclear temperature $T$, defined as
\begin{equation}\label{NT}
\frac{1}{T} = \frac{d}{d E} \ln \rho (E).
\end{equation}
However, ${\mathcal{J}}$, in reality, is not a constant, but a changing quantity with $E$. Generally, ${\mathcal{J}}$ is smaller than ${\mathcal{J}}_{\rm rigid}$ as it approaches the highest excitation, and at low excitation, deviations in either direction can be expected. 

\begin{figure} \centering
\includegraphics[width=3.3in]{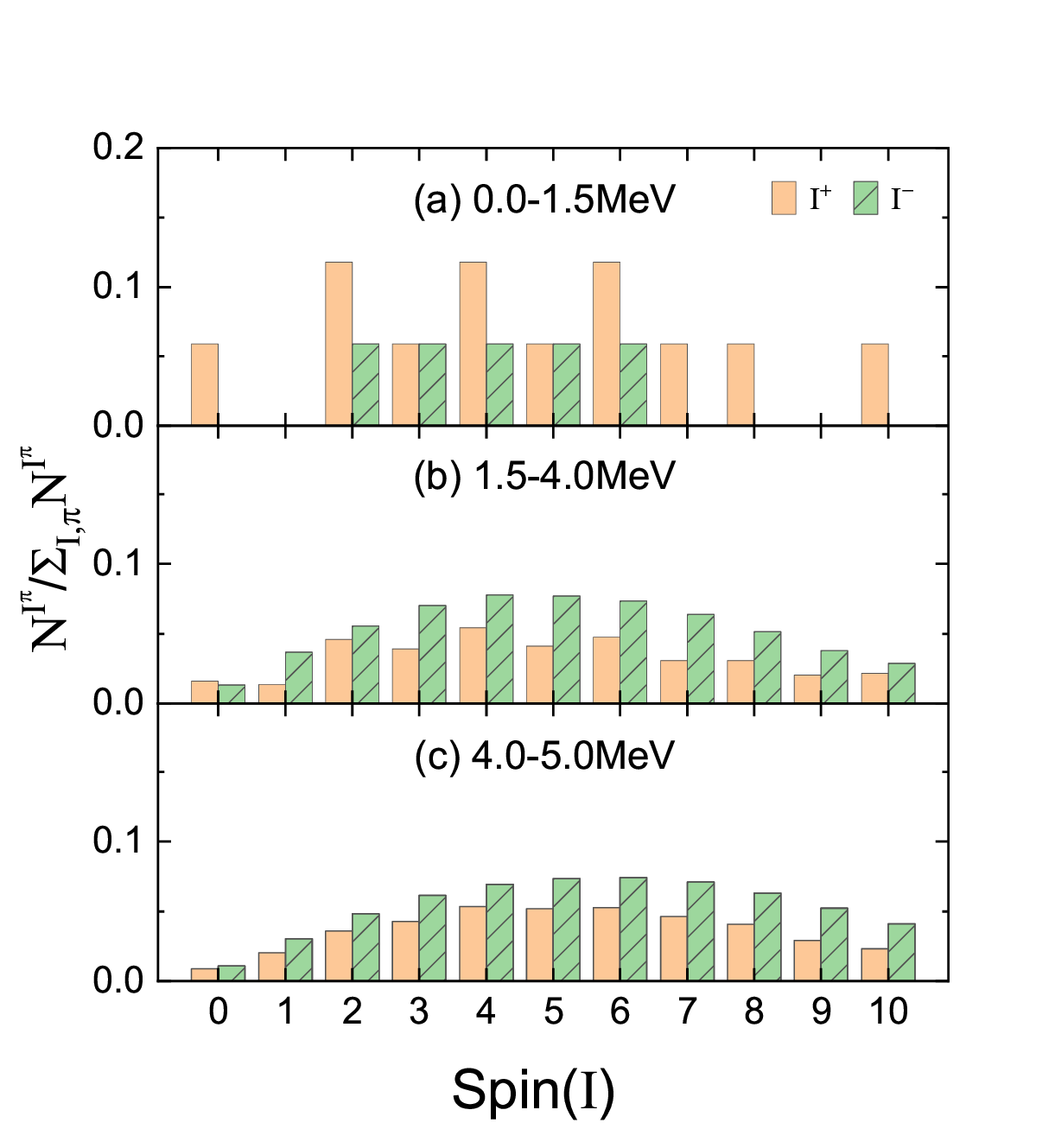}
\caption{\label{fig:spins}(Color online) 
Histogram of integrated level density in different regimes as functions of angular momentum (spin), decomposed into even and odd parities, and normalized to one over each respective integrated energy range.}
\end{figure}

The calculated states by the PSM are eigenstates of spin and parity. Therefore, we have precise information on spin-parity dependence for all calculated levels, as well as on internal transitions connecting them. The latter feature of the PSM provides a basis for us to study $\gamma$ strength functions. To see the spin dependence, we group the calculated total NLD results in $^{164}$Dy according to spin over the three different energy regimes discussed in Sec. V-B. The accumulated levels in each of the three energy regimes are normalized to one, decomposed into even and odd parities, and drawn as functions of spin in the form of histograms as shown in the three segmented plots of Fig. \ref{fig:spins}. Therefore, what we have shown in the three different histogram plots of Fig. \ref{fig:spins}, are the relative distribution probability of NLD as a function of spin-parity for definite energy ranges. In the following, we discuss the characteristic features of these plots.


\begin{description}
   \item[Fig. \ref{fig:spins}(a):] It shows the result of the first energy regime, i.e. the collective regime (0 - 1.5 MeV). For the collective states in even-even nuclei, one has a good understanding of the structure and knows well the spin and parity for each level. In the current $^{164}$Dy example, only 17 known levels fall in this regime, among which 5 are of odd parity. 
   \item[Figs. \ref{fig:spins}(b):] The plot shows the spin-parity dependence of NLD for the pair-breaking regime (1.5 - 4.0 MeV). This regime covers the entire pair-breaking process starting from the breaking of the very first nucleon-pair till the boundary energy at which almost all the neutron- and proton-pairs, no matter what orbitals they occupy, are broken. We can view this regime, in contrast to the collective one in (a) and the chaotic one in (c), as a {\it transitional} regime where the actual content of breaking pairs, thus the wavefunctions, are changing with excitation. Thus physically, this is the most interesting regime. Overall in Fig. \ref{fig:spins}(b), one sees that there are more odd-parity states than even-parity ones, consistent with what we have seen in Fig. \ref{fig:parity}. Furthermore, for even-parity states in Fig. \ref{fig:spins}(b), irregular spin distribution is seen, indicating a pronounced structure-dependence, while for odd parity, the distribution shows a tendency to become bell-shaped. Especially, for the even-parity states, there is an interesting odd-even effect in spin for which an even-spin distribution is clearly larger than either of its neighboring odd-spin ones. Interstingly, similar odd-even effect in spin was found before by Alhassid {\it et al.} in Ref. \cite{Alhassid2007} from the SMMC calculation for the light even-even nucleus $^{56}$Fe.
   \item[Fig. \ref{fig:spins}(c):] In this plot with energy range 4.0 - 5.0 MeV where the total level density approaches $10^4$ (see plots in Figs. \ref{fig:level} and \ref{fig:parity}), we observe nearly perfect Gaussians in the calculated spin distribution for both parities. The peak of the Gaussian for even parity is centered around $I=4, 5$ and 6, while that for odd parity is at $I= 5$ and 6, which clearly imply distinct $\sigma$ values in Eq. (\ref{sigma}). The odd-even effect in spin seen in Fig. \ref{fig:spins}(b) seem to beome much weaker here. This energy regime, and excitations higher than 5 MeV, consists of multi-qp states {\it all} from broken pairs. One can expect Gaussian-type patterns in the spin distribution, and therefore, applying the spin-distribution using Eq. (\ref{spin-distribution}) begins to make sense.
\end{description}

\begin{figure*} \centering
\includegraphics[width=6.3in]{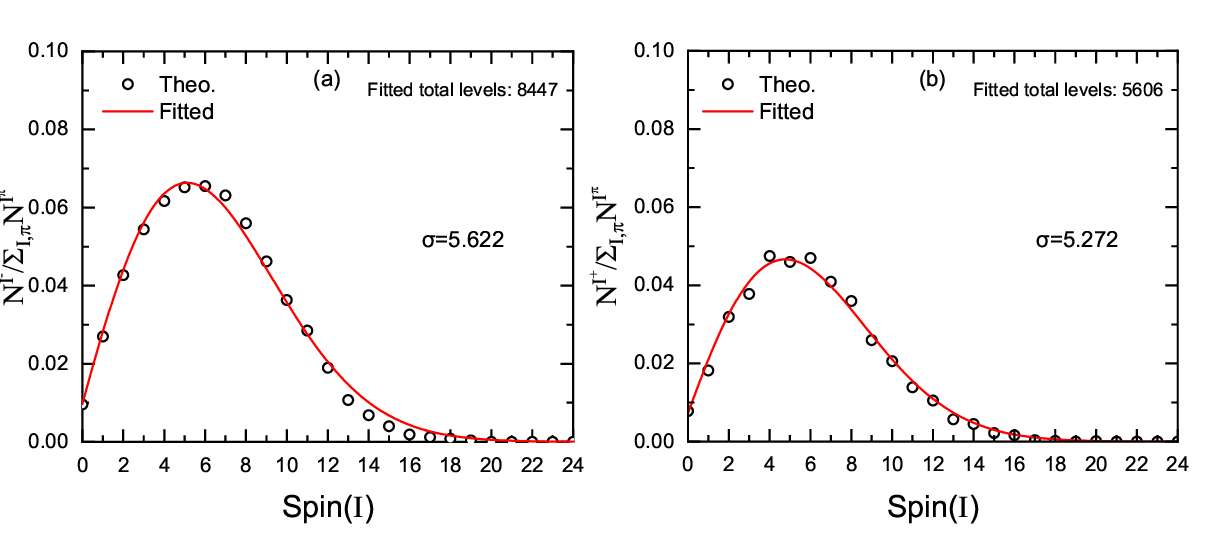}
\caption{\label{fig:gaussians}(Color online) 
Least-squares fitting with the Ericson's formula of Eq. (\ref{spin-distribution}) for the spin-distribution in the levels with $4.0-5.0$ MeV, for (a) odd-parity, and (b) even-parity.}
\end{figure*}

From Fig. \ref{fig:spins}(c), it can be seen that for the multi-qp regime (i.e. the chaotic regime), our shell-model calculation indicates Gaussian-like spin distribution as suggested earlier by Ericson \cite{Ericson1959,Ericson1960}.
Moreover, since the present result comes out from a spectroscopic calculation, it can help in deepening our understanding of the spin-dependent NLD at least in two aspects. Firstly, it provides a detailed description of the low-energy collective and the pair-breaking regimes where the spin-distribution is generally irregular, and thus Eq. (\ref{spin-distribution}) is invalid. Secondly, for the higher-energy multi-qp regime where Eq. (\ref{spin-distribution}) may apply, the value of the dispersion $\sigma$, which determines the shapes of the Gaussian and is closely related to nuclear structure properties, can now be obtained using shell-model calculations without the need of approximation in Eq. (\ref{sigma}). The determination of the $\sigma$-parameter has been discussed by some authors (see, for example, Refs. \cite{Voinov2007,Grimes2016}).

It is straightforward to perform a numerical fitting for the calculated spin-distribution in Fig. \ref{fig:spins}(c) to the Ericson formula in Eq. (\ref{spin-distribution}), and find our theoretical $\sigma$. We adopt the least-squares method by using the MATLAB {\it lsqcurvefit}. In Fig. \ref{fig:gaussians}, we show, respectively in (a) and (b) for odd- and even-parity, the fitting results from the levels in Fig. \ref{fig:spins}(c) within the energy range of $4.0-5.0$ MeV. Note that in order to include the long tail in the high-spin part of the Gaussian, we extend the projection calculation up to $I=24\hbar$. Therefore, as compared to Fig. \ref{fig:spins}(c), Fig. \ref{fig:gaussians} include more levels: 8447 levels with odd-parity and 5606 levels with even-parity. Consequently, we obtain $\sigma=5.622$ and 5.272 for odd- and even-parity, respectively. We remark that $\sigma$ alone absorbs all the nuclear structure information containing in the spin distribution, and calculation by using microscopic models is therefore of great interest. 

The 2007 work of Alhassid {\it et al.} \cite{Alhassid2007} on spin distribution has significant overlap with our current results, despite of totally different concept of the model. Originally in their SMMC approach, thermal averages are taken over all possible states of a given nucleus, and thus the computed level densities are those summed over all possible spin states. The authors introduced a spin-projection method that enables to calculate
thermal observables at  given spins. Their discussion \cite{Alhassid2007}  used the complete $(pf + g_{9/2})$ shell for the iron-region nuclei. As they clearly showed, for even-even nuclei and at low excitation energies, they observed a similar odd-even staggering in the spin dependence of level densities, which, according to the authors, was related to the pairing effect. In the same work \cite{Alhassid2007}, $\sigma$-values were explicitly extracted from the calculation as functions of excitation energy.

A close inspection in the plots in Fig. \ref{fig:spins} indicates that the odd-parity levels follow the statistical Gaussian spin-distribution in lower energies, even in the pair-breaking regime (see Fig. \ref{fig:spins}(b)). This is clearly different from the even-parity levels of the same energy region where an odd-even staggering in the spin dependence is clearly evident. Another interesting observation in Ref.~\cite{Alhassid2007} is that in the case of odd-mass and odd-odd nuclei, such a spin-staggering effect is much suppressed. Notice that in the odd-parity levels in even-even nuclei and in odd-mass and odd-odd nuclei, there are always qp states that do not participate in the pair correlation, thus effectively reducing the pairing of the system.

\section{Comparison with the Shell Model Monte Carlo approach}\label{Compare-SMMC}

The study of NLD has been a very old subject since the time of Hans Bethe. Many models have been developed for theoretical calculations. In particular, the current PSM work may have significant overlaps with the Shell Model Monte Carlo (SMMC) approach \cite{smmc_2}, which has been successfully applied to the NLD study \cite{Alhassid2008,Alhassid2013,Alhassid2015}. Although we have already mentioned some of the SMMC results throughout the paper, we try to summarize the similarities and differences as well as advantages and disadvantages of the SMMC as compared to the PSM in this section.

For a quantitative, microscopic description of physical observables, one in principle solves the many-body eigenvalue equation, $\hat H\left|\Psi\right> = E \left|\Psi\right>$, to obtain all solutions in the Hilbert space. Here $\left|\Psi\right>$ is the $many$-$body$ wavefunction, containing all necessary information on the participating particles with the interactions among them. Since exact solutions for the present problem are not possible, one is compelled to seek  approximations \cite{Sun}. The PSM, as introduced in the early part of the present paper, adopts a large single-particle space but a restricted many-body configuration space in terms of pair-breakings. Within the truncated configuration space, it is possible to solve the many-body eigenvalue problem, and obtain all eigenvalues and eigenstates. NLD's with definite spin/parity can then be directly constructed from the solution. 

The SMMC \cite{smmc_2} treats the many-body problem differently. Instead of solving the eigenvalue problem directly, it deals with the imaginary-time many-body evolution operator $\exp(-\beta \hat H)$, with $\beta$ being a $c$-number, which is reduced to a coherent superposition of one-body evolutions in fluctuating one-body fields. The resultant path integral is evaluated stochastically. In contrast to the PSM, the SMMC does not result in a complete solution to the many-body problem in the sense of giving all eigenvalues and eigenstates of $\hat H$. However, it can still obtain very useful information when the expectation value of  observables in the grand canonical ensemble is  calculated \cite{smmc_2}. For the NLD problem, the calculated total NLD contains those with all possible spins and both parities. NLD's with definite spin/parity can be obtained by spin- and parity-projection from the total NLD \cite{Alhassid2007,Alhassid1997}. 

NLD is a statistical quantity by definition. Hence a SMMC description has naturally advantages. This is particularly true for the high-energy region near the neutron resonance energy, where the NLD reaches one million per MeV interval in the $^{164}$Dy example. With such dense levels, character for individual levels becomes unimportant. This is the place that we call the chaotic regime in the present paper. For this energy region, exact diagonalization calculation by the PSM requires heavy computation effort. It is not practical to carry out a systematical NLD calculation with the exact method for the high-energy region. For the current cut-off energy at 4.5 MeV (see Fig. 2), it is already very time-consuming with our usual workstation. On the other hand, for NLD in this regime, it does not necessarily require precise spectroscopic properties. This is one obvious disadvantage of the PSM.

One advantage of the PSM approach for the NLD calculation may lie in the lower-energy region, especially in the phase-transition region that we call the pair-breaking regime. This energy region is characterized by rich structure changes and the interplay between the collective and quasi-particle motions. A good understanding of the levels in this regime has a potential importance that it determines the energy from which the linear behavior of NLD starts. This is of  practical help for the Oslo experiment to serve as the anchor point at the low-energy side, as the existing discrete data in the nuclear data base are far from being complete, and therefore, unrealistic.

Both approximations, PSM and SMMC, are valid nuclear many-body methods, but in finding many-body solutions, they are based on completely different concepts, namely, quantum mechanics versus statistics. It is interesting to find out connections between the two models, when the two methods are applied to studying the same physical quantity and if both can describe experiments successfully. This question deserves an investigation.

\section{SUMMARY AND FUTURE prospective}\label{Summary}

Nuclear level density is a basic property of atomic nuclei and is a crucial ingredient in nuclear reaction theories. For the research field of nuclear astrophysics, nuclear level density, together with several other nuclear structure quantities, serve as important inputs to the study of the rapid neutron capture process (r process), which is believed to be responsible for about half of the production of the elements heavier than iron \cite{Kajino2019}. All these inputs are  nuclear structure quantities, among which, the nuclear mass is a sole ground state property. The rest, such as internal electromagnetic transitions, $\beta$ decay rates, and neutron capture rates, involve excited nuclear states. In principle, a nuclear shell model is able to calculate {\it all} these quantities in a consistent way.

For example, for a quantitative, microscopic description of nuclear level density, one should solve the exact many-body eigenvalue problem, $\hat H\left|\Psi\right> = E \left|\Psi\right>$, and obtain all energy levels in the Hilbert space. However, this has turned out to be an impossible task for mid-mass and heavy nuclei if the discussion is confined in the conventional shell model. One has to develop novel shell-model methods by applying many-body techniques. The present work takes a step towards this goal. We extend the Projected Shell Model to study, in a systematical and consistent way, nuclear level density in the present paper and $\gamma$ strength function and other related problems in forthcoming papers. It is aimed at providing nuclear structure information to all these quantities, so that one may discuss these statistical quantities together with concrete nuclear structure. 

The current article is the first one in our planned series of publications. As this is the first application of the PSM in the calculation of statistical quantities, we have introduced the basic nuclear structure for excited nuclear states and emphasized the unique process of pair breaking under nuclear rotation and thermal excitation. We discussed the gradual evolution of pair breaking in rotating nuclei which usually occurs over a wide range of excitations. Taking this phenomenon into consideration, we have outlined the multi-qp structure of the PSM. The essence of the PSM is that the model is constructed based on the deformed basis with the angular-momentum projection technique. Thus the deformed single-particle states and the formation of multi-qp configurations enter into the discussion for level-density calculations. 

Throughout the paper, the nuclear structure effect appears in the discussion. We have shown that the consequences of pair breaking sensitively influence the behavior of NLD at different excitations. The particularly interesting excitation range is 1.5 - 4.0 MeV in the $^{164}$Dy example, which we call the pair-breaking regime. The pair-breaking effect in this regime strongly depends on the deformed single-particle states. We have found a new step structure in NLD beyond 2.0 MeV of excitation, which physically corresponds to a simultaneous breaking of neutron- and a proton-pair. Once this prediction is confirmed, it may amend the general understanding of the NLD curve in low excitations. We have also shown that for $^{164}$Dy, different from the usual thought \cite{Ericson1960}, the total NLD can not be simply divided into two equal groups of even and odd parity, not even approximately. This is again attributed to the structure effect of the deformed single-particle orbitals. 
Moreover, we have studied the distribution of NLD with different spins and found that for low-energy regions before the consummation of pair breaking, spin distribution patterns are generally irregular and structure-sensitive. With increasing energy beyond 4 MeV in our $^{164}$Dy example, a Gaussian-like distribution emerges, consistent with what Ericson suggested \cite{Ericson1960,Ericson1959}. This provides a possibility that by fitting the Ericson distribution with our shell-model result, one can determine the unknown dispersion $\sigma$ in the Ericson formula, which is a structure-dependent quantity.

The present shell-model calculation is limited in lower excitations (up to $E \sim$ 5 MeV and $\rho < 10^5$ in the $^{164}$Dy example). In principle, the extension of calculation to higher energy regions is possible but requires heavy computational effort. 
On the other hand, the numerical effort can be greatly reduced if we just use the full set of multi-qp configurations in (\ref{shell_config}) to count levels, without carrying out configuration mixing calculation in very large matrices. Preliminary results show that we can obtain an NLD curve qualitatively similar to the one presented in Fig. (\ref{fig:level}), and can easily extend the calculation up to the neutron separation energy to compare the result with the NLD from the neutron-resonance spacing data. This work will be presented in a separate publication \cite{Dutta2023}.   

In odd-mass nuclei, one last nucleon is blocked from the pair formation. Usually, a few (and up to about ten) deformed 1-qp states exist around the Fermi levels in an odd-mass nucleus. Therefore, depending on the deformed single-particle structure, the level density of an odd-mass nucleus is on average a few to ten times higher than those of their neighboring even-even counterparts at a same excitation energy. Guttormsen {\it et al.} found in Ref. \cite{Guttormsen2000} that the level densities for the $^{161}$Dy and $^{171}$Yb isotopes are about five times higher than that for the neighboring $^{162}$Dy and $^{172}$Yb isotopes. However, they pointed out that the conclusion works well only for excitation energies between 3.5 and 7 MeV, which are about the energies of the multi-qp regime suggested in the present work. We suspect that for the low-energy range below 3.5 MeV in odd-mass nuclei, individual single-particle structure will determine the NLD pattern. In Ref. \cite{Guttormsen2001}, the authors, by examining about 280 nuclei, argued interestingly that for these odd-mass nuclei, the entropy scales with the number of particles that are not coupled in Cooper pairs. Detailed shell-model study is in progress, and the odd-mass NLD will be the discussion focus in our forthcoming paper.

\begin{acknowledgments}
One of us (YS) acknowledges the first conversation with M. Wiedeking and A. V. Voinov at NS2022 in Berkeley. We thank these gentlemen, and also M. Mumpower and W. Misch for their interests in applying the present results to nuclear astrophysics. This work is supported by the National Natural Science Foundation of China (Grant Nos. 12235003, 12275225, and U1932206).
\end{acknowledgments}


\end{document}